# Unidirectional Amplification in the Frozen Mode Regime Enabled by a Nonlinear Defect

S. LANDERS[1], W. TUXBURY[1], I. VITEBSKIY[2], T. KOTTOS[1]

[1]Wave Transport in Complex Systems Lab, Department of Physics, Wesleyan University, Middletown CT-06459, USA
[2]Air Force Research Laboratory, Sensors Directorate, Wright-Patterson Air Force Base, OH- 45433, USA

**ABSTRACT**

A stationary inflection point (SIP) is a spectral singularity of the Bloch dispersion relation $\omega(k)$ of a periodic structure where the first and the second derivatives of $\omega$ with respect to $k$ vanish. An SIP is associated with a third order exceptional point degeneracy in the spectrum of the unit-cell transfer matrix, where there is a collapse of one propagating and two evanescent Bloch modes. At the SIP frequency, the incident wave can be efficiently converted into the frozen mode with greatly enhanced amplitude and vanishing group velocity. This can be very attractive for applications, including light amplification. Due to its non-resonant nature, the frozen mode regime (FMR) has fundamental advantages over common cavity resonances. Here, we propose a novel scheme for FMR-based unidirectional amplifiers by leveraging a tailored amplification/attenuation mechanism and a single nonlinear defect. The defect breaks the directional symmetry of the periodic structure and enables nonlinearity-related unidirectional amplification/ attenuation in the vicinity of the SIP frequency. We demonstrate the robustness of the amplification mechanism to local impurities and parasitic nonlinearity.

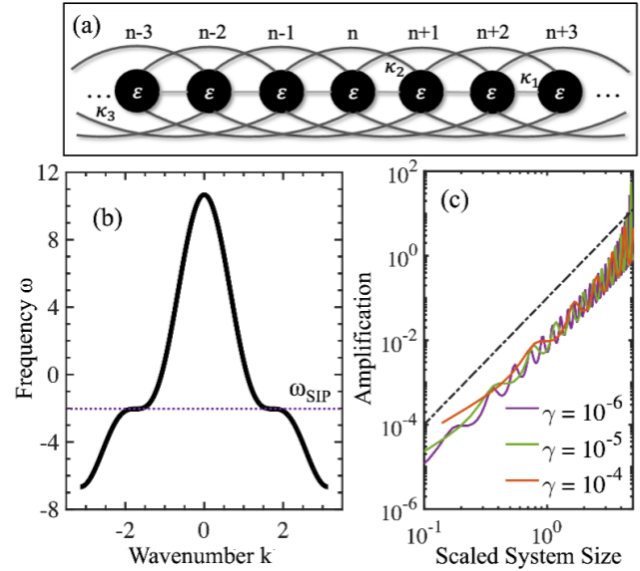

**Fig. 1. Coupled Mode Theory Model.** (a) Tight-binding schematic of coupled mode theory (CMT) model consisting of first, second and third order long-range coupling. (b) Dispersion relation exhibiting symmetric SIPs at $k_{SIP} \approx \pm 1.738$ and $\omega_{SIP} \approx -2.037$. (c) Amplification, $T + R - 1$, at frequency $\omega_{SIP}$ plotted versus the re-scaled system size $L\gamma^{1/3}$, for various uniform gain strengths indicated in the legend. The black dashed line shows the cubic dependency of the transmitted signal.

**Introduction** Physical systems described by non-Hermitian operators have attracted growing attention over the previous decade across the subfields of wave physics [1–3]. In particular, unique types of spectral degeneracies known as exceptional point degeneracies (EPDs) can emerge, where both eigenvalues and their corresponding eigenvectors coalesce [4]. Exceptional point degeneracies have undoubtedly spurred many new developments including unidirectional invisibility, hypersensitive sensing, enhanced power emission, chiral state transfer, and more [3]. In all these cases, however, the formation of EPDs was confined to the resonance modes of the system i.e., purely outgoing solutions of the relevant wave equation.

In this letter we investigate a class of non-resonant third-order EPDs (EPD-3) occurring in the spectrum of the unit-cell transfer matrix of a periodic waveguide. An EPD-3 is reflected in the Bloch dispersion relation (BDR) $\omega(k)$, which develops a stationary inflection point (SIP) at the associated frequency. Specifically, in the proximity of the EPD, the BDR takes the form [5], $\nu \sim \delta k^3$, where $\nu \equiv \omega - \omega_{SIP}$ and $\delta k \equiv k - k_{SIP}$ are the frequency and wavenumber detuning away from the SIP. Remarkably, at the SIP frequency, the incident wave can be converted into the *frozen mode* with little or even no reflection (see, for example [5], and references therein). The frozen mode inside the waveguide has a hugely enhanced amplitude and vanishing group velocity. In proximity to the SIP frequency, the continuity of the wave amplitude at the waveguide interface is achieved by a destructive interference between three diverging Bloch components: one propagating, with vanishing group velocity, and two evanescent, with vanishing imaginary parts of the respective Bloch wavenumbers [5]. The enhanced amplitude of the frozen mode and its efficient coupling with the input wave makes the frozen mode regime (FMR) an attractive alternative to cavity resonances. An additional advantage of the FMR is its robustness to structural imperfections and losses, which is due to its non-resonant nature [5–9]. Here, we harness these unique properties of the FMR to develop a design protocol for SIP-based unidirectional, imperfection-tolerant amplifiers.

The objective of this Letter is to establish a generic scheme that can be used to design an FMR-based unidirectional amplifier. First, we introduce a generic coupled mode theory (CMT) model which supports an SIP in the BDR. A local nonlinearity divides the structure into two asymmetric domains, each consisting of loss or gain elements. The asymmetric non-Hermiticity imposes a directional dependence of the interaction strength of the wave at the nonlinear defect, which is enhanced at the SIP frequency due to its anomalous scaling properties of amplification/attenuation lengths. Finally, we assess the robustness of the SIP-based asymmetric transport to the presence of local impurities and globally distributed (parasitic) Kerr nonlinearity. The design principles emerging from this work can serve as a springboard for even more SIP-based technologies.

**Model Description** - Generic features associated with the FMR (e.g., enhanced field intensity, anomalous absorption length) are independent of the specific model or platform that gives rise to the SIP [6,7,10,11]. In this respect, we introduce a minimal model based on CMT [12–14] to serve as a numerical testbed to affirm the proposed schema. The physical description of the underlying model consists of a one-dimensional periodic array of identical oscillators with long-range coupling between them. A schematic of this model is displayed in figure 1a. For this model to exhibit an SIP, its defining parameters require tunability up to the third-order coupling.

The dynamics of the underlying system are described by a tri-block diagonal Hamiltonian operator $H$ such that, in the steady-state, the equations of motion of the $n^{th}$ unit cell satisfy,

$$\omega\psi_n = \varepsilon\psi_n + \sum_{l=1}^{3} \kappa_l(\psi_{n+l} + \psi_{n-l}) \quad (1)$$

The dispersion relation, $\omega(k)$, associated with the model of Eq. (1) can be immediately obtained from Bloch's theorem, $\psi_{n\pm l} = \psi_n e^{\pm ikal}$, where $a = 1$ is the spacing between the resonators, normalized to unity. The associated dispersion relation is,

$$\omega(k) = \varepsilon + 2\sum_{l=1}^{3} \kappa_l \cos(lk) \quad (2)$$

For the sake of convenience, the energy reference is defined by setting the natural frequency of the independent oscillators to zero, $\varepsilon = 0$. Additionally, the energy scale is defined by setting the third nearest-neighbor coupling equal to one, $\kappa_3 = 1$. Meanwhile, Eq. (2) displays an SIP when the nearest and next nearest-neighbor coupling values are $\kappa_1 = 10/3$ and $\kappa_2 = 1$. The BDR exhibiting symmetric SIPs at $k_{SIP} \approx \pm 1.738$ and $\omega_{SIP} \approx -2.037$ is displayed in figure 1b.

The periodic system is turned to a scattering setting by coupling leads to the first and last sites of the truncated lattice, consisting of $L$ unit cells. The dynamics are governed by the $L \times L$ Hamiltonian matrix $H_0$ of the isolated system, with $\varepsilon$ along the main diagonal and $\kappa_l$ along the $\pm l^{th}$ diagonals for $l = 1, 2, 3$. Each lead is modeled by a semi-infinite chain of oscillators with nearest-neighbor coupling $\kappa_L = \kappa_1$ and frequency $\varepsilon_L = \varepsilon$. The dispersion relation of the leads is given by $\omega = \varepsilon + 2\kappa_L \cos k$ and results in a self-energy correction to the effective Hamiltonian that describes the scattering setup, $H_{eff} = H_0 + \kappa_L e^{ik} WW^T$, where the matrix $WW^T$ projects the correction to the sites where the leads are attached. Effects of the cavity-induced Fabry-Perot resonances can be mitigated by including a gradual (logistic function with logistic growth $k_B$) tapering of the coupling parameters on the right side of the structure over $L_B$ unit-cells, which we refer to as the buffer, to match those in the leads (details are included in the caption of figure 2), while the signatures of the SIP remains unaffected due to its non-resonant nature. Our description of the leads, while numerically practical, also introduces artificial pathologies associated with their band edges which will be neglected in our further considerations.

For a linear system, transport can be computed via a scattering matrix analysis where the internal dynamics are described in the frequency domain by the equations of motion, $\omega|\psi\rangle = H_{eff}|\psi\rangle + iD|s^+\rangle$, supplemented with the continuity relation $|s^-\rangle = D^T|\psi\rangle - |s^+\rangle$, where $|\psi\rangle$ describes the scattering field inside the structure, $|s^\pm\rangle$ encodes the amplitudes of the incoming/outgoing propagating modes in the leads and $D = \Im m\left[\sqrt{2\kappa_L e^{ik}}\right] W$ is an $L \times M$ matrix called the coupling matrix. Combining these results the scattering matrix $|s^-\rangle = \hat{S}|s^+\rangle; \hat{S} = -\mathbb{I}_M + iD^T GD$, where $\mathbb{I}_M$ is the $M \times M$ identity matrix and $G \equiv [\omega \mathbb{I}_L - H_{eff}]^{-1}$ is the Green's function, from which the transport can be computed.

A unique SIP-related phenomena is an anomalous scaling in its absorption length $\xi_\gamma \propto \gamma^{-1/3}$ when the system is supplemented by uniform loss $\gamma$ [15]. Here, we have also confirmed an analogous anomalous scaling in the amplification by plotting transmittance (absent the buffer) as a function of the re-scaled system size $\lambda \propto L\gamma^{1/3}$ (prior to lasing) in figure 1c, where $\gamma$ indicates now a uniform gain. Therefore, both rapid absorption $A \sim L^3$ in the presence of loss and a rapid amplification $T \sim L^3$ in the presence of gain is observed at the SIP frequency. These will serve as the primary design mechanisms, together with an embedded nonlinear defect, that will enhance unidirectional amplification.

**Unidirectional Amplifier** - We utilize the anomalous attenuation/amplification behavior of the SIPs, together with a tailored gain/loss profile and an asymmetrically positioned nonlinearity, to achieve unidirectional amplification. The forward solution of the scattering problem, in the presence of a single on-site nonlinearity, can be directly computed by neglecting excitation of higher-order harmonics. For a nonlinearity of the generic form $f = f\left(\left|\psi_{n_0}\right|^2\right)$, we express the equations of motion in the frequency domain as $\omega|\psi\rangle = H_{eff}|\psi\rangle + \psi_{n_0} f|n_0\rangle + iD|s^+\rangle$. We introduce the $L - 1 \times L$ filtering matrix $F$ with the defining properties $F|n_0\rangle = 0$, $F^T F + |n_0\rangle\langle n_0| = \mathbb{I}_L$ and $FF^T = \mathbb{I}_{L-1}$. Specifically, the elements of $F$ are $[F]_{ij} = \begin{cases} \delta_{i,j}, & i < n_0 \\ \delta_{i,j-1}, & i \geq n_0 \end{cases}$. Inserting $\mathbb{I}_L$ we get

$$(\omega - H_{eff})(F^T F + |n_0\rangle\langle n_0|)|\psi\rangle = \psi_{n_0} f|n_0\rangle + iD|s^+\rangle \quad (3)$$

Multiplying Eq. (3) by $F$, after re-arrangement, we can express the linear components of the field as $F|\psi\rangle = \psi_{n_0} \tilde{G} F H_{eff}|n_0\rangle + i\tilde{G}FD|s^+\rangle$, where $\tilde{G} = [F(\omega\mathbb{I}_L - H_{eff})F^T]^{-1}$. Instead, application of $\langle n_0|$ to Eq. (3) leads to $(\omega - \varepsilon_{n_0})\psi_{n_0} = \langle n_0|H_{eff}F^T F|\psi\rangle + \psi_{n_0} f + i\langle n_0|D|s^+\rangle$, where $\varepsilon_{n_0} \equiv \langle n_0|H_{eff}|n_0\rangle$. Inserting the former into the latter and re-arranging, we arrive at an equation for the field amplitude at the nonlinear site,

$$\beta_0 = (\beta_1 + f)\psi_{n_0} \quad (4)$$

where $\beta_0 = -i\langle n_0|(H_{eff}F^T \tilde{G} F + \mathbb{I}_L)D|s^+\rangle$ and $\beta_1 = \langle n_0|H_{eff}F^T \tilde{G} F H_{eff}|n_0\rangle - (\omega - \varepsilon_{n_0})$. Computing the modulus

squared of Eq. (4) results in a cubic polynomial in $|\psi_{n_0}|^2$ that is solved using Cardano's formula [8]. A physical requirement is that the field intensity must be real and positive, $|\psi_{n_0}|^2 \in \mathbb{R}^+$. Once the physically admissible solution is identified, it can be used to evaluate $f\left(|\psi_{n_0}|^2\right)$, which enables use of the linear scattering formalism (see above) for the given input wavefront $|s^+\rangle$ and frequency $\omega$.

Conceptually, the operational mechanism of the unidirectional amplifier can be understood from figure 2a. The design is composed of a relatively short lossy region on the left (green) and a longer stretch of gain to the right (red), separated by a Hermitian nonlinear defect (black strip). The loss and gain are modeled by adding imaginary on-site potentials $-i\gamma_L$ and $+i\gamma_G$, respectively, to the diagonal elements of the effective Hamiltonian (details are provided in the caption of figure 2). The nonlinear defect is modeled as a Kerr-type nonlinearity at the site $n_0$, $f\left(|\psi_{n_0}|^2\right) = \chi|\psi_{n_0}|^2$, where $\chi \equiv 1$ is the nonlinear coefficient, and $\psi_{n_0}$ is the scattering field amplitude of the nonlinear site.

With the conceptual and theoretical framework outlined, consider a wave incident at the SIP frequency from the left. First, the slow wave will be promptly attenuated, enabling the current to pass the nonlinear site unimpeded. Then, the amplitude is dramatically intensified by the gain and the amplified signal is transmitted with diminished back-reflection, abated by the buffer. In contrast, a wave at the same frequency incident from the right will undergo intense amplification and subsequently be primarily reflected due to the impedance mismatch induced by the nonlinear defect. Importantly, the gigantic intensification and drastic attenuation effects, enabling unidirectional amplification, rely on the reduced amplification and absorption lengths unique to the SIP.

In figure 2b, we present the transmission as a function of the incident power. At extremely high or low input intensities, transmission is reciprocal, while in an intermediate regime, where the input power is $P_{in} = \langle s^+|s^+\rangle \approx 2.465 \times 10^{-4}$, unidirectional amplification is achieved with $T_L \approx 2.266$ and $T_R \approx 0.88$. The inset displays the real and imaginary parts of the recursively computed, self-consistent eigenfrequencies corresponding to the generalized eigenvalue problem, $H_{eff}(\omega_n)|\psi_n\rangle = \omega_n|\psi_n\rangle$ for $\chi = 0$. Since $\Im m(\omega_n) < 0$ for low input power and the outgoing flux remains bounded as the input is ramped, the system is guaranteed to operate below the lasing threshold.

**Robustness to Local Impurities and Intrinsic Nonlinearity** - The robust character of SIP phenomena to inevitable imperfections, e.g., due to fabrication tolerances or material properties of the platform, has been demonstrated for passive, linear systems [6,9,15,16]. Here, we show that this robustness to local impurities is preserved even in the presence of gain and the nonlinear defect. Moreover, we verify that the SIP-enabled features of the device are maintained even when global, intrinsic nonlinearities are involved.

*Local Impurities* - To analyze the resilience of the unidirectional amplification to impurities, we randomly selected 10% of the resonators and modify their resonance frequency by an amount $\delta\varepsilon$, chosen from a uniform distribution $\delta\varepsilon \in [-\Delta/2, \Delta/2]$. We have checked that the system remains below its lasing threshold. The evaluated transmittances are presented in figure 3a as a function of the frequency at the incident power which results in the optimal transmission contrast indicated in figure 2b with the vertical double-arrow. Impurity strengths $\Delta$ of 0.1% and 1% were considered and both preserve the unidirectional amplification close to that of the ideal scenario. Notice that at frequencies away from the SIP the transmission is reciprocal.

*Intrinsic Nonlinearity* - The field enhancement in proximity to the SIP frequency is a key mechanism underlying the directional amplification. However, this growth in the field intensity may also trigger unintended secondary (parasitic) nonlinear effects. To validate the proposed model with respect to this parasitic mechanism, we use the so-called backward map [17–19] to compute transmission in the presence of globally distributed nonlinearity, which is modeled as an intensity-dependent frequency shift among all previously linear $n \neq n_0$ sites of the structure, $\eta|\psi_n|^2$, where $\eta$ is the Kerr coefficient. The ratio of the Kerr coefficient at the defect site to that of the rest of the sites is taken to be $\eta/\chi = 10^{-3}$ to emulate the contrast between the nonlinear properties of some common dielectric materials [20].

The basic ingredient of the backward map is the transfer matrix of the unit cell, where now the unit cell must be formulated to consist of three sites to avoid non-local interactions (i.e., beyond adjacent unit cells). At the same time, the unit cells of the leads must also be modified to accommodate the same dimensionality as inside the structure. Therefore, two additional leads on both sides are coupled to the remaining sites in the first and last unit cells, otherwise identical to the already existing leads. As in the previous section, we include a buffer on the right side over $N_b$ unit cells, but where now $\kappa_1, \kappa_2 \to 0$ and $\kappa_3 \to \kappa_L$. In such case, the equations of motion for the $n^{th}$ unit cell are $\omega|\psi_n\rangle = V_n^\dagger |\psi_{n-1}\rangle + U_n|\psi_n\rangle + V_{n+1}|\psi_{n+1}\rangle$, where $|\psi_n\rangle$ is a $3 \times 1$ vector whose components describe the complex field amplitudes on the sites of the $n^{th}$ unit cell, $V_{n+1}$ is a $3 \times 3$ matrix encoding the couplings from sites of the $n^{th}$ unit cell to that of the unit cell $n+1$, and $U_n$ is a $3 \times 3$ matrix containing the frequencies and coupling elements within the $n^{th}$ unit cell, and also carries the nonlinear dependencies in that unit cell. Therefore, we can write the field in the $n-1$ unit cell in terms of the field in the $n$ and $n+1$ unit cells, $|\psi_{n-1}\rangle = \left(V_n^\dagger\right)^{-1}(\omega \mathbb{I}_3 - U_n)|\psi_n\rangle - \left(V_n^\dagger\right)^{-1} V_{n+1}|\psi_{n+1}\rangle$. The backward map is performed by first prescribing the transmitted wave in the leads $|\psi_{N+l}\rangle = e^{ikl}|t\rangle$; $l \geq 0$, which is taken to be the optimal transmission eigenstate of the linear structure, and then using $|\psi_{N+1}\rangle$ and $|\psi_N\rangle$, which also determines $U_N$, in the formula above to evaluate $|\psi_{N-1}\rangle$. Recursively evaluating $U_n$ and $|\psi_{n-1}\rangle$ from $|\psi_n\rangle$ and $|\psi_{n+1}\rangle$ computes the entire scattering field and, in particular, the wave in the input leads $|\psi_{-l}\rangle = e^{ikla}|i\rangle + e^{-ikla}|r\rangle$; $l \geq 0$, which can be used to decouple the incident $|i\rangle$ and reflected wavefronts $|r\rangle$, providing the transmission $T = \frac{\langle t|t\rangle}{\langle i|i\rangle}$ as a function of the input frequency $\omega$ and incident power $P_{in} = \langle i|i\rangle$.

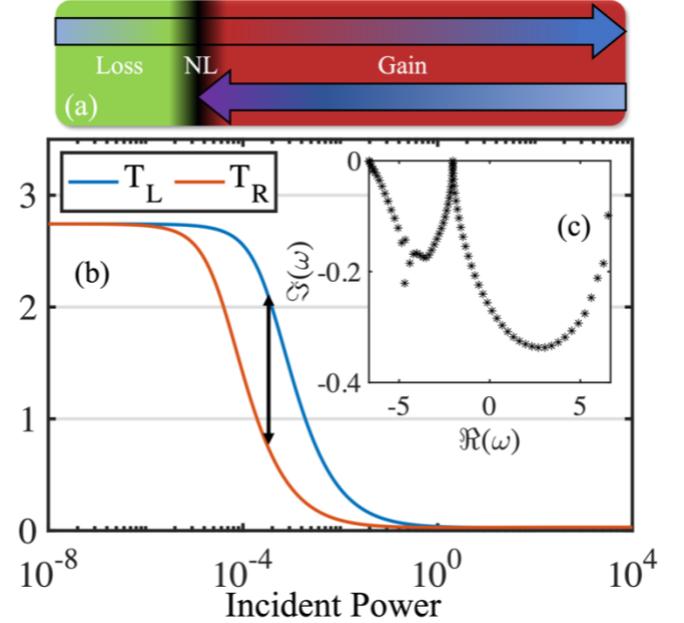

**Fig. 2. Unidirectional Amplifier.** (a) Schematic protocol for SIP-based unidirectional amplification. Loss ($\gamma_L = 1.4 \times 10^{-4}$) is positioned to the left of a single nonlinear site ($\chi = 1$, $n_0 = 20$), while gain ($\gamma_G = 2.4 \times 10^{-4}$) is placed on the right side. The structure consists of $L = 100$ unit-cells and is constructed with a buffer of length $L_B = 20$ with logistic growth rate $k_B = 0.75$. A wave incident from the left at the SIP frequency initially undergoes rapid attenuation, enabling it to pass the nonlinearity and be amplified in the gain portion, while incident from the right, the wave amplitude is bolstered by the gain and predominantly reflected at the nonlinear site. (b) Directional transmission as a function of the incident power. The power at which the transmission asymmetry is maximum is indicated by the black vertical arrows. (c) Eigenvalues of the Effective Hamiltonian, indicating the linear system operates below lasing threshold.

The results of the backward map are presented in figure 3b. The qualitative behavior of the ideal structure is reproduced as, in proximity to the SIP at the optimal input power $P_{in} \approx 5 \times 10^{-5}$, transmission incident from the left is amplified, $T_L \approx 1.463$, and transmission incident from the right is suppressed, $T_R \approx 0.286$.

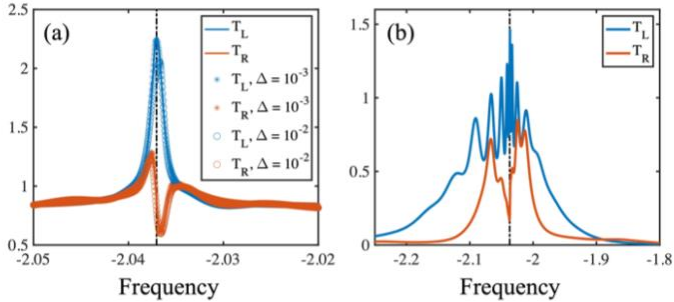

**Fig. 3. Robustness to Impurity and Nonlinearity.** (a) Directional transmission as a function of frequency at the incident power that results the optimal transmission contrast, indicated by black arrows in figure 2b, for the system with impurities among 10% of its lattice with the impurity strength ranging from 0 (ideal) to 1% ($\Delta = 10^{-2}$) indicated by the different markers in the legend. (b) Directional transmission computed via the backward map in a system with $N = 33$, $N_b = 7$ and $k_B = 2.25$. The nonlinearity strength throughout the system is $10^{-3}$ compared with that of the nonlinear defect. Apparent amplification reduction is not a consequence of the added nonlinearity, but rather due to the altered boundary conditions to accommodate the backward map. In either case, the SIP-based design demonstrates adequate resilience to fabrication imperfection as well as self-induced disorder as a nonlinear consequence of the dramatic SIP field intensity enhancement.

The backward map provides a qualitative verification of the robustness of the proposed SIP-based unidirectional amplifier to intrinsic nonlinearities as can be seen from the preserved contrast between the left-going and right-going wave transmission in proximity to the SIP frequency.

**Discussion -** In this Letter, we introduced design protocols for unidirectional amplifiers based on the frozen mode interaction with integrated nonlinear defect. Using CMT modeling, we have shown how anomalously fast absorption and amplification in the FMR, together with the dramatically enhanced frozen mode amplitude, can be utilized for the design of magnetic-free, unidirectional amplifiers. We expect the proposed design schema to steer research in non-reciprocal technology towards utilization of the unique features of the FMR.

**Disclosures**
The authors declare that they have no competing interests.

**Funding**
KBR, DE-SC0024223, AFOSR LRIR 24RYCOR008

**References**
1. R. El-Ganainy, K. G. Makris, M. Khajavikhan, Z. H. Musslimani, S. Rotter, and D. N. Christodoulides, "Non-Hermitian physics and PT symmetry," Nat Phys **14**, 11–19 (2018).
2. C. M. Bender, P. E. Dorey, C. Dunning, A. Fring, D. W. Hook, H. F. Jones, S. Kuzhel, G. Lévai, and R. Tateo, *PT Symmetry in Quantum and Classical Systems* (World Scientific (Europe), 2018).
3. M.-A. Miri and A. Alú, "Exceptional points in optics and photonics," Science (1979) **363**, (2019).
4. T. Kato, *Perturbation Theory for Linear Operators* (Springer, 1995), Vol. 132.
5. A. Figotin and I. Vitebskiy, "Slow wave phenomena in photonic crystals," Laser Photon Rev **5**, 201–213 (2011).
6. W. Tuxbury, R. Kononchuk, and T. Kottos, "Non-resonant exceptional points as enablers of noise-resilient sensors," Commun Phys **5**, 210 (2022).
7. A. Figotin and I. Vitebskiy, "Slow light in photonic crystals," Waves in Random and Complex Media **16**, 293–382 (2006).
8. C.-Z. Wang, R. Kononchuk, U. Kuhl, and T. Kottos, "Loss-Induced Violation of the Fundamental Transmittance-Asymmetry Bound in Nonlinear Complex Wave Systems," Phys Rev Lett **131**, 123801 (2023).
9. H. Li, I. Vitebskiy, and T. Kottos, "Frozen mode regime in finite periodic structures," Phys Rev B **96**, 180301 (2017).
10. N. Gutman, C. M. de Sterke, A. A. Sukhorukov, and L. C. Botten, "Slow and frozen light in optical waveguides with multiple gratings: Degenerate band edges and stationary inflection points," Phys. Rev. A **85**, 033804 (2012).
11. K. Zamir-Abramovich, N. Furman, A. Herrero-Parareda, F. Capolino, and J. Scheuer, "Low-threshold lasing with a stationary inflection point in a three-coupled-waveguide structure," Phys Rev A (Coll Park) **108**, 63504 (2023).
12. H. A. Haus, *Waves and Fields in Optoelectronics* (Prentice Hall, 1984).
13. W. Suh, Z. Wang, and S. Fan, "Temporal coupled-mode theory and the presence of non-orthogonal modes in lossless multimode cavities," IEEE J Quantum Electron **40**, 1511–1518 (2004).
14. S. Landers, A. Kurnosov, W. Tuxbury, I. Vitebskiy, and T. Kottos, "Nonlinear wavepacket dynamics in proximity to a stationary inflection point," Phys Rev B **109**, 24312 (2024).
15. W. Tuxbury, L. J. Fernandez-Alcazar, I. Vitebskiy, and T. Kottos, "Scaling theory of absorption in the frozen mode regime," Opt. Lett. **46**, 3053–3056 (2021).
16. Z. M. Gan, H. Li, and T. Kottos, "Effects of disorder in frozen-mode light," Opt Lett **44**, 2891–2894 (2019).
17. D. Hennig and G. P. Tsironis, "Wave transmission in nonlinear lattices," Phys Rep **307**, 333–432 (1999).
18. P. Markoš and C. M. Soukoulis, *Wave Propagation* (Princeton University Press, 2008).
19. E. Lidorikis, K. Busch, Q. Li, C. T. Chan, and C. M. Soukoulis, "Optical nonlinear response of a single nonlinear dielectric layer sandwiched between two linear dielectric structures," Phys Rev B **56**, 15090–15099 (1997).
20. M. Karlsson, "Nonlinear Propagation of Optical Pulses and Beams," (1994).